# RANDOM IRREGULAR BLOCK-HIERARCHICAL NETWORKS: ALGORITHMS FOR COMPUTATION OF MAIN PROPERTIES

Svetlana Avetisyan    Mikayel Samvelyan*    Martun Karapetyan

Yerevan State University

**Abstract**

In this paper, the class of random irregular block-hierarchical networks is defined and algorithms for generation and calculation of network properties are described. The algorithms presented for this class of networks are more efficient than known algorithms both in computation time and memory usage and can be used to analyze topological properties of such networks. The algorithms are implemented in the system created by the authors for the study of topological and statistical properties of random networks.

## 1. Introduction

Block-hierarchical networks have been of particular interest in recent years as they turned out to present a convenient model of the spatial structure of complex biomolecules, such as DNA and proteins [1]. An active study of various complex networks gives relevance to the development of a system that can simulate models of random graphs of different types and conduct an effective and comprehensive analysis of their topological and statistical properties [2]. The main properties of a random network include node degree distribution, node-to-node distance distribution, the distribution of the clustering coefficients of the nodes, cycle length distribution, length distribution of the connected subnetworks, network diameter, etc.

In this paper, the class of random irregular block-hierarchical networks is defined and algorithms for generation and calculation of network properties are described. Block-hierarchical networks can be represented as a hierarchy of clusters (subnets) connected to each other. In regular case, which are discussed in [3,4], the number of nested clusters, and hence the number of nodes within the clusters of the same level are identical. In case of an irregular block-hierarchical network, the number of nested clusters may range from 1 to $p$ where $p$ specifies the maximum possible number of nested clusters. Algorithms for the class of irregular block-hierarchical networks, which coincide with those for regular networks, are not listed here (refer to [3]). All the results are used in the implementation of "Random Networks Explorer" [4] which enables to carry out a study of statistical properties of random networks. The results of the analysis for the degree distribution of regular and irregular block-hierarchical networks are provided as an illustration.

## 2. An irregular block-hierarchical network

An irregular block-hierarchical network is a generalized case of a regular block-hierarchical network [3]. Let's define a the former first. A regular block-hierarchical network $G_{p,\Gamma} \in \mathscr{R}_{p,\Gamma}$ is defined by 2 parameters: $p$ - branching index and $\Gamma$ - the number of levels in the network, $p > 1, \Gamma \geq 0$. The number of network nodes $\{x_1, \dots, x_N\}$, where $N = p^\Gamma$ and partition into clusters is precisely defined by these parameters, which determines the structure of the nodes in a regular block-hierarchical network. The network is constructed in levels. At each new level $\gamma$, $0 \leq \gamma \leq \Gamma$, new clusters are formed through merging $p$ of the clusters that have already been constructed at the previous level. This results in the formation of new connections between the nodes of the network, as all nodes of a cluster are connected to every node of a different cluster, if the two are connected to each other on the same level of hierarchy. The node-link tree represents partition into clusters, while the bit sequences that mark all the vertices of the node-link tree determine connection of the clusters, i.e. the presence of links in the network. $M_\gamma^{(i)}$ represents $i$-th cluster on the level $\gamma$, $1 \leq i \leq n_\gamma$, where $n_\gamma$ is the number of clusters on the level $\gamma$. $M_\gamma$ represents a set of clusters on the level $\gamma$,

$$M_\gamma = \{M_\gamma^{(1)}, M_\gamma^{(2)}, \ldots, M_\gamma^{(n_\gamma)}\}$$

V($M_\gamma^{(i)}$) defines nodes within cluster $M_\gamma^{(i)}$. All networks in the class $\mathfrak{R}_{p,\Gamma}$ have the same number of nodes and the same partition into clusters and differ in the choice of connected pairs of nested clusters only.

For the irregular case, $p$ specifies the maximum number of partitions into clusters; the number of sub-clusters $\leq p$ for each cluster. $p$ and $\Gamma$ do not clearly determine partitioning into clusters and do not set the number of nodes, as is the case with regular networks. In case of any level of $1 \leq \gamma \leq \Gamma$,

$$\bigcup_{i=1}^{n_\gamma} V\left(M_\gamma^{(i)}\right) = V\left(M_\Gamma^{(1)}\right) = \{x_1, \ldots, x_N\}, \qquad N \leq p^\Gamma$$

Let $Count(\gamma, i)$ denote the number of clusters nested in the cluster $M_\gamma^{(i)}$, $Count(0, i) = 0$. Partition into clusters on the level $\gamma > 0$ is determined by the following set:

$$Branch(\gamma) = \{Count(\gamma, i) | 1 \leq i \leq n_\gamma\}.$$

$Branch(0) = \emptyset$, and will not be examined further. It's obvious that $|Branch(\gamma - 1)| = \sum_{i=1}^{n_\gamma} Count(\gamma, i)$, $1 \leq \gamma \leq \Gamma$. Partition of the entire network is determined by the following set:

$$Branch = \{Branch(\gamma) | 1 \leq \gamma \leq \Gamma\}$$

For this partition of $Branch$, a class of irregular block-hierarchical networks is defined $G \in \mathfrak{R}_{Branch}$. Networks of this class have the same structure of partition of nodes into clusters and only differ in connection among network nodes. The class of regular block-hierarchical networks $\mathfrak{R}_{p,\Gamma}$ coincides with that of irregular case $\mathfrak{R}_{Branch}$, if for each $i$ and $\gamma$, $1 \leq i \leq n_\gamma$, $1 \leq \gamma \leq \Gamma$, $Count(\gamma, i) = p$. In that case, $N = p^\Gamma$.

*The node-link tree is used for generating and storing the network.* The node-link tree of the network $G$ is set in the form of a $(p + 1)$-tree that satisfies the following conditions:

- At each level $\gamma$, $0 \leq \gamma \leq \Gamma$ there are $n_\gamma$ vertices $t_\gamma^{(i)}$, $1 \leq i \leq n_\gamma$. The number of sub-trees at the vertex of $t_\gamma^{(i)}$ is $Count(\gamma, i)$, $Count(0, i) = 0$. The total number of vertices of the node-link tree is equal to the number of clusters of the network $G$.

- Each cluster $M_\gamma^{(i)}$ of the network $G$ is associated with the sub-tree whose root is the vertex of the node-link tree $t_\gamma^{(i)}$. The nodes in the cluster $M_\gamma^{(i)}$ correspond to the leaves (the end vertices) of the respective sub-tree.

- The vertex of the node-link tree $t_\gamma^{(i)}$, $1 \leq \gamma \leq \Gamma$ is marked by a sequence of zeroes and ones $bitmap(\gamma, i)$, with a length of $\frac{k*(k-1)}{2}$, $k = Count(\gamma, i)$. The bit sequence $bitmap(\gamma, i)$ describes the connection between the clusters on the level of $\gamma - 1$, nested in the cluster $M_\gamma^{(i)}$. Let's label $bitmap(\gamma, i)$ a node-link vector of the vertex $t_\gamma^{(i)}$ of the node-link tree.

The node-link tree of an irregular block-hierarchical network $G$ is defined by the structure of partition into clusters:

$Branch = \{Branch(\gamma) | 1 \leq \gamma \leq \Gamma\}$, where $Branch(\gamma) = \{Count(\gamma, i) | 1 \leq i \leq n_\gamma\}$,

and a set of $Bitmap$ - set of node-link vectors:

$Bitmap = \{Bitmap(\gamma) | 1 \leq \gamma \leq \Gamma\}$, where $Bitmap(\gamma) = \{bitmap(\gamma, i) | 1 \leq i \leq n_\gamma\}$.

The block-hierarchical network $G \in \Re_{Branch}$ is determined precisely by a set of $Branch$ and $Bitmap$. The sub-tree corresponding to the cluster will be denoted similar to the cluster as $M_\gamma^{(i)}$. As an illustration in Figure 1, the network $G$ and the corresponding node-link tree are presented, wherein

$Branch = \{\{3, 4, 2\}, \{3\}\}, Bitmap = \{\{<011>, <100110>, <1>\}, \{<100>\}\}$,

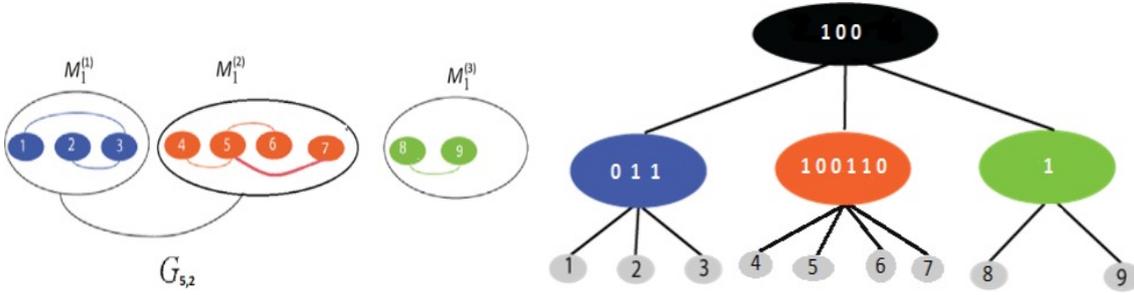

Fig.1. Node-link tree of network $G \in \Re_{Branch}$.

## 3. Algorithms for generation of irregular block-hierarchical networks

Generation of network $G \in \Re_{Branch}$ is creation of a corresponding node-link tree, i.e. generation of structure of partition of nodes into clusters.

$Branch = \{Branch(\gamma) | 1 \leq \gamma \leq \Gamma\}$, where $Branch(\gamma) = \{Count(\gamma, i) | 1 \leq i \leq n_\gamma\}$,

and generation of multiple node-link vectors

$Bitmap = \{Bitmap(\gamma) | 1 \leq \gamma \leq \Gamma\}$, where $Bitmap(\gamma) = \{bitmap(\gamma, i) | 1 \leq i \leq n_\gamma\}$,

Generation of $Branch$ is possible by setting the maximum branching index $p$ and either the number of elements of the network $N$, or the number of levels $\Gamma$. An algorithm for both cases is described below.

To generate node-link vector $bitmap(\gamma, i)$, the probability of connection $\omega = k^{-\mu}$, $k = |V(M_\gamma^{(i)})|$, $\mu > 0$ is determined. $\mu$ specifies the generated network density, while $k$ – the number of nodes in the cluster $M_\gamma^{(i)}$.

### 3.1. Generation of the structure of partition into clusters when setting maximum branching index $p$ and number of nodes $N$

Generation of the structure of partition into clusters $Branch$ is carried out by levels from the bottom up, i.e. from the leaves of the tree to the root. Clusters on the level $\gamma$ are constructed, followed by a random and equiprobable combination of some clusters on the level $\gamma$ with those on the level $\gamma + 1$. The number of levels $\Gamma$ of an irregular block-hierarchical network is determined in the process of network generation $\lceil N/p \rceil \leq \Gamma < \infty$.

Input: $N, p$
Output: $\Gamma$ – the level of the constructed tree, $Branch$ – the structure of partition into clusters.
1. $Branch \coloneqq \emptyset, n \coloneqq N, \gamma \coloneqq 0$
2. If $n = 1$, then $\Gamma \coloneqq \gamma$ and complete algorithm
3. $\gamma \coloneqq \gamma + 1, i \coloneqq 1$
4. Generate a random integer $l$ from 1 to p
5. If $l > n$, then $l \coloneqq n$
6. $Count(\gamma, i) \coloneqq l, n \coloneqq n - l, Branch(\gamma) \coloneqq Branch(\gamma) \cup \{l\}, i \coloneqq i + 1$
7. Repeat Steps 4-6 until $n \neq 0$
8. $n \coloneqq |Branch(\gamma)|, Branch \coloneqq Branch \cup \{ Branch(\gamma) \}$, return to Step 2

### 3.2. Generation of structure of partition into clusters for a given $p$ and number of hierarchical levels $\Gamma$

Generation is carried out by the levels of hierarchy from top to bottom, i.e. the root is generated first, while the leaves of the tree, which correspond to the nodes of the network, are generated last. The number of nodes in an irregular block-hierarchical network is determined in the process of network generation.

Input: $p, \Gamma$
Output: $N$ – number of network nodes, $Branch$ – structure of partition into clusters.
1. $\gamma \coloneqq \Gamma, Branch \coloneqq \emptyset, n \coloneqq 1, i \coloneqq 1, m \coloneqq 0$
2. If $\gamma = 0$, then complete the algorithm with the output $N \coloneqq n$
3. Generate a random integer $l$ from 1 to p
4. $Count(\gamma, i) \coloneqq l, Branch(\gamma) \coloneqq Branch(\gamma) \cup \{l\}, i \coloneqq i + 1, m \coloneqq m + l$
5. Repeat Steps 3-4 n times
6. $n \coloneqq m, Branch \coloneqq \{Branch(\gamma)\} \cup Branch, \gamma \coloneqq \gamma - 1, i \coloneqq 1, m \coloneqq 0$
7. Return to Step 2

### 3.3. Generation of node-link vectors

Generation of node-link vectors $Bitmap$ is carried out on already created node-link tree $Branch$.

Input: $\mu > 0, \Gamma, Branch$
Output: $Bitmap$ – multiple node-link vectors.
1. $Bitmap \coloneqq \emptyset, \gamma \coloneqq 0$
2. If $\gamma = \Gamma$, then complete the algorithm
3. $i \coloneqq 1, \gamma \coloneqq \gamma + 1$
4. Determine $k \coloneqq |V(M_\gamma^{(i)})|$ - number of nodes in cluster $M_\gamma^{(i)}$
5. Generate bit sequence $bitmap(\gamma, i)$ with the length of $l * (l - 1)/2$, where $l = Count(\gamma, i)$. Probability of connection $\omega \coloneqq k^{-\mu}$ is used when generating
6. $Bitmap(\gamma) \coloneqq Bitmap(\gamma) \cup \{bitmap(\gamma, i)\}, i \coloneqq i + 1$
7. Repeat Steps 5-6 $n$ times
8. $Bitmap \coloneqq Bitmap \cup \{Bitmap(\gamma)\}$
9. Return to Step 2

Let's estimate the number of levels in the generated network. The average value of a random variable $Count(\gamma, i)$ is $(p+1)/2$ which means that the average number of network levels is $\Gamma_{aver} = \lceil \log_{(p+1)/2} N \rceil$ when using generation algorithm 3.1. This particular estimate is practiced to assess the complexity of algorithms for calculation of the main properties of an irregular block-hierarchical network.

## 4. Algorithms for calculation of the main properties of an irregular block-hierarchical network

The node-link tree determines the structure of storage in a block-hierarchical network, and generation of a random network $G$ with nodes $N$ is reduced to generation of a set of *Branch* and *Bitmap*. All the algorithms being developed use this storage structure. The calculation of network properties is carried out along the node-link tree which provides high efficiency both in memory usage and computation time. In this chapter, algorithms for calculating the degree of a network node, the distance between two network nodes, the number of node-links in a network, the number of connected subgraphs of a given length, the number of cycles with length of 3 and the number of cycles with length of 4 are described. The assessment of complexity is presented for comparison with the classical algorithms.

A definition similar to the one in [3] is introduced. $S\left(M_\gamma^{(i)}\right), 1 \leq \gamma \leq \Gamma$ denotes a set of clusters on the level $\gamma - 1$, nested in the cluster $M_\gamma^{(i)}$.

$$S(M_\gamma^{(i)}) = \{S_1(M_\gamma^{(i)}), S_2(M_\gamma^{(i)}), \ldots, S_k(M_\gamma^{(i)})\}, k = Count(\gamma, i), \quad 1 \leq i \leq n_\gamma,$$

where $S_n(M_\gamma^{(i)})$, $1 \leq n \leq k$ – nested cluster $n$. Let's define the function $\psi_{\gamma,i}(n, s)$ of direct connection of two sub-clusters $S_n(M_\gamma^{(i)}), S_s(M_\gamma^{(i)})$ as follows:

$$\psi_{\gamma,i}(n, s) = \begin{cases} 1, & \text{if clusters } S_n(M_\gamma^{(i)}) \text{ and } S_s(M_\gamma^{(i)}) \text{ are connected directly} \\ 0, & \text{otherwise} \end{cases}$$

Sub-clusters $S_n(M_\gamma^{(i)})$ and $S_s(M_\gamma^{(i)})$ are connected directly if the bit describing the relation between them is 1 in the node-link vector $bitmap(\gamma, i)$.

### 4.1. Calculation of the degree of a node in the irregular block-hierarchical network

Suppose $G \in \mathcal{R}_{Branch}$ is a block-hierarchical network, $x \in V(G)$. Let's consider the set of clusters on the level $\gamma$ $\{M_\gamma^{(1)}, M_\gamma^{(2)}, \ldots, M_\gamma^{(n_\gamma)}\}$. The function $v(x, \gamma)$ determines the number of clusters on the level $\gamma$, containing the node $x$. In this case, $M_\gamma^{(v(x,\gamma))}$ is the only cluster on the level $\gamma$, containing the node $x$. For convenience, let's denote it $M_\gamma^{(x)}$.

*Claim 1.* Let's suppose $M_\gamma^{(x)}$ is a cluster on the level $\gamma$, in that case

$$degree\left(x, M_\gamma^{(x)},\right) = \sum_{i=1}^{\gamma} \sum_{j=1}^{Count(i,v(x,i))} \left(\psi_{i,v(x,i)}(v(x, i-1), j) * |V(S_j(M_i^{(v(x,i))}))|\right)$$

where $degree\left(x, M_\gamma^{(x)}\right)$ is the degree of the node $x$ in the cluster $M_\gamma^{(x)}$, $1 \leq i \leq \gamma$.

*Proof.* To calculate the degree of the node $x$ in the cluster $M_\gamma^{(x)}$ it's sufficient to examine the path from the leaf $x$ to the vertex $t_\gamma^{(n)}$ on the node-link tree and at each level of $i$, $1 \leq i \leq \gamma$ and to count the number of direct connections in the cluster $M_{i-1}^{(x)}$ along the node-link vector. Claim 1 is proven.

Network $G$ will have $degree(x, G) = degree\left(x, M_{\Gamma}^{(1)}\right)$

*Assessment of the complexity of the algorithm.* To calculate $|V(S_j(M_i^{(v(x,i))}))|$ it is necessary to traverse all the vertices of the node-link tree of the respective cluster $M_{i-1}^{(j)}$ other than Level 0. The maximum number of vertices of the node-link tree $p$ that are not leaves is $\sum_{k=0}^{i-1} p^k$. To calculate $degree(x, G)$ it's necessary to carry out the steps of $\sum_{i=1}^{\Gamma}((p-1) * \sum_{k=0}^{i-1} p^k)$. The branching index $p$ is a constant and is determined prior to the calculation. When $\Gamma_{aver} = \lceil log_{(p+1)/2} N \rceil$ the time complexity of the algorithm is $O(N)$.

### 4.2. Distance between two nodes

*Claim 2.* Let's suppose $d(x, y)$ is the distance between two connected vertices of the network. In that case $d(x, y) \in \{1, 2\}$, if p = 2 and $d(x, y) \in \{1, 2, ..., p-1\}$, if $p \geq 3$.

The proof and algorithm for computation of the distance between two nodes is similar to the regular case [3].

*Assessment of the complexity of the algorithm.* As with the regular case, it's sufficient to go up the node-link tree from leaves of the respective nodes to the vertex of the tree that corresponds to the cluster on the lowest level which contains the given nodes. When $\Gamma_{aver} = \lceil log_{(p+1)/2} N \rceil$ the time complexity of the algorithm is $O(logN)$.

### 4.3. Number of edges in the irregular block-hierarchical network

*Claim 3.* Let's suppose $G \in \mathcal{R}_{Branch}$, $E(M_\gamma^{(n)})$ – number of edges in the cluster $M_\gamma^{(n)}$, $1 \leq n \leq n_\gamma$, in that case

$$|E(M_\gamma^{(n)})| = \sum_{k=1}^{c} \left|E\left(S_k\left(M_\gamma^{(n)}\right)\right)\right| + \sum_{i=1}^{c-1}\left(\sum_{j=i+1}^{c}\left(\psi_{\gamma,n}(i,j) * |V(S_i(M_\gamma^{(n)}))| * |V(S_j(M_\gamma^{(n)}))|\right)\right),$$

where $c = Count(\gamma, n)$, $E(M_0^{(i)}) = \emptyset$, $1 \leq i \leq N$.
Network $G$ will have $|E(G)| = \left|E(M_\Gamma^{(1)})\right|$.

*Assessment of the complexity of the algorithm.* First of all, note that, for the calculation of the number of vertices of a given cluster, it's not necessary to traverse all the vertices of the node-link tree of the respective cluster, it's sufficient to sum up the number of vertices of the sub-clusters that have already been counted by the algorithm $|V(M_\gamma^{(n)})| = \sum_{k=1}^{c} |V(S_k(M_\gamma^{(n)}))|$. Hence it follows that a maximum of $p^2$ steps is required for the calculation of the addend. Operating time of the algorithm for calculation of $|E(M_\gamma^{(n)})|$ satisfies the following recurrence relation:

$$\begin{cases} F(\gamma) = p * F(\gamma - 1) + p^2, & \gamma > 0 \\ F(\gamma) = 0, & \gamma = 0, \end{cases}$$

where $F(\gamma)$ denotes the complexity $|E(M_\gamma^{(n)})|$, $1 \leq n \leq n_\gamma$. Solving it, we get $F(\gamma) = p^2 * (p^\gamma - 1)/(p - 1)$. For level $\gamma = \Gamma$ we have $O(p^2 * (p^\Gamma - 1)/(p - 1))$. When $\Gamma_{aver} = \lceil log_{(p+1)/2} N \rceil$ the time complexity of the algorithm is $O(N)$.

### 4.4. Number of cycles with length of 3 in the irregular block-hierarchical network

*Claim 4.* Let's suppose $G \in \mathcal{R}_{Branch}$ and $Cycles(M_\gamma^{(n)}, 3)$ is the number of cycles with length of 3 between cluster nodes $M_\gamma^{(n)}$, in that case

$$Cycles(M_\gamma^{(n)}, 3) = \sum_{i=1}^{c} Cycles\left(S_i(M_\gamma^{(n)}), 3\right) +$$

$$\sum_{\substack{i,j=1\ldots c \\ i \neq j}} \left(\psi_{\gamma,n}(i,j) * \left|E\left(S_i(M_\gamma^{(n)})\right)\right| * |V(S_j(M_\gamma^{(n)}))|\right) +$$

$$\sum_{i=1}^{c-2} \sum_{j=i+1}^{c-1} \sum_{k=j+1}^{c} \left(\psi_{\gamma,n}(i,j) * \psi_{\gamma,n}(j,k) * \psi_{\gamma,n}(k,i) * |V(S_i(M_\gamma^{(n)}))| * |V(S_j(M_\gamma^{(n)}))|\right.$$

$$\left. * |V(S_k(M_\gamma^{(n)}))|\right)$$

where $c = Count(\gamma, n)$, $Cycles(M_0^{(i)}, 3) = 0$, $1 \leq i \leq N$,
Network $G$ will have $Cycles(G, 3) = Cycles(M_\Gamma^{(1)}, 3)$.

*Assessment of the complexity of the algorithm.* Operating time of the algorithm for calculation of $Cycles(M_\gamma^{(n)}, 3)$ satisfies the following recurrence relation:

$$\begin{cases} F(\gamma) = p * F(\gamma-1) + p^3 + p^2, & \gamma > 0 \\ F(\gamma) = 0, & \gamma = 0, \end{cases}$$

where $F(\gamma)$ denotes the complexity $Cycles(M_\gamma^{(n)}, 3)$, $1 \leq n \leq n_\gamma$. Solving it, we get $F(\gamma) = p^2 * (p+1) * (p^\gamma - 1)/(p-1)$. When $\Gamma_{aver} = \lceil log_{(p+1)/2} N \rceil$ the time complexity of the algorithm is $O(N)$.

### 4.5. Number of different cycles with length of 3 passing through the same node

Let $Cycles_x(M_\gamma^{(x)}, 3)$ denote the number of different cycles with length of 3 passing through the same node $x$ in the cluster $M_\gamma^{(x)}$

*Claim 5.* Let's suppose $G \in \mathcal{R}_{Branch}$ is a block-hierarchical network, in that case

$$Cycles_x(M_\gamma^{(x)}, 3) = Cycles_x(M_{\gamma-1}^{(x)}, 3) + \sum_{i=1}^{c} \left(\psi_{\gamma,v(x,\gamma)}(v(x,\gamma-1), j) * \left|E(S_j(M_\gamma^{(x)}))\right|\right) +$$

$$\sum_{j=1}^{c} \left(\psi_{\gamma,v(x,\gamma)}(v(x,\gamma-1), j) * degree(M_{\gamma-1}^{(x)}, x) * \left|V\left(S_j(M_\gamma^{(x)})\right)\right|\right) +$$

$$\sum_{i=1}^{c-1} \sum_{j=i+1}^{c} \left(\psi_{\gamma,v(x,\gamma)}(v(x,\gamma-1), i) * \psi_{\gamma,v(x,\gamma)}(i,j) * \psi_{\gamma,v(x,\gamma)}(j, v(x,\gamma-1)) * \left|V\left(S_i(M_\gamma^{(x)})\right)\right|\right.$$

$$\left. * \left|V\left(S_j(M_\gamma^{(x)})\right)\right|\right)$$

*Assessment of the complexity of the algorithm.* As in the previous cases, the number of cycles with length of 3, passing through the same node in the network cluster, can be calculated by traversing the node-link tree once. When $\Gamma_{aver} = \lceil log_{(p+1)/2} N \rceil$ the time complexity of the algorithm is $O(N)$.

### 4.6. Number of cycles with length of 4 in the irregular block-hierarchical network

Similar to Claim 3, an algorithm is developed for calculating the number of cycles with length of 4 in an irregular block-hierarchical network. When $\Gamma = \lceil log_{(p+1)/2}N \rceil$, the time complexity of the algorithm is $O(N)$.

***Results of the experiment.*** The given algorithms were tested in the system under development, "Random Networks Explorer". The results of the two experiments are as follows:
1. Calculation of cycles with length of 3 and 4 for regular and irregular block-hierarchical networks (Table 1).
2. Charts for the degree distribution of nodes for regular and irregular block-hierarchical networks (Fig. 2).

The results were obtained on an assembly of 100 copies, for networks with branching parameter $p = 3$ and the number of vertices $N = 3^9 = 19683$.

| N | p | μ | Cycles with length of 3 | | Cycles with length of 4 | |
|---|---|---|---|---|---|---|
| | | | regular network | irregular network | regular network | irregular network |
| 19683 | 3 | 0.1 | $\approx 2 * 10^{11}$ | $\approx 2.5 * 10^{11}$ | $\approx 2 * 10^{15}$ | $\approx 2.3 * 10^{15}$ |
| 19683 | 3 | 0.3 | $\approx 2.5 * 10^9$ | $\approx 5 * 10^9$ | $\approx 1.2 * 10^{14}$ | $\approx 6 * 10^{13}$ |
| 19683 | 3 | 0.5 | $\approx 8 * 10^7$ | $\approx 2.5 * 10^8$ | $\approx 2 * 10^{13}$ | $\approx 3.5 * 10^{12}$ |
| 19683 | 3 | 0.8 | $\approx 4 * 10^5$ | $\approx 10^7$ | $\approx 4 * 10^9$ | $\approx 4 * 10^{10}$ |

*Table 1. Comparison of cycles with length of 3 and 4 for regular and irregular block-hierarchical networks where = 3, N = 19683 on an assembly of 100 copies*

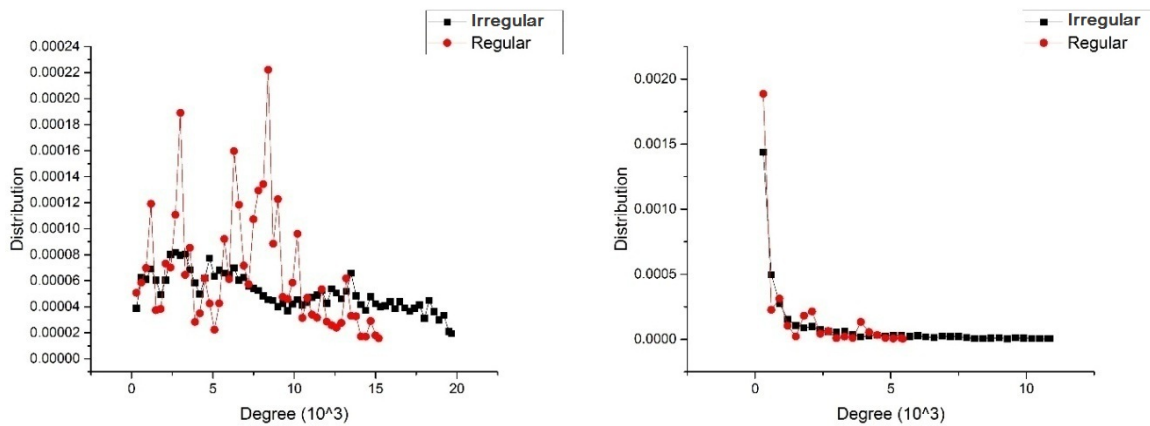

*Fig. 2. Comparison of the degree distribution for irregular (round) and regular (square) networks where $N = 19683, p = 3, \mu = 0.1$ (on the right), $\mu = 0.3$ (on the left) on an assembly of 100 copies*